\documentclass[sn-mathphys]{sn-jnl}

\jyear{2022}%
\usepackage{amsmath}
\begin{document}

\title[Kramers-Moyal analysis of IMF fluctuations at sub-ion scales]{Kramers-Moyal analysis of interplanetary magnetic field fluctuations at sub-ion scales}

\author*[1]{\fnm{Simone} \sur{Benella}}\email{simone.benella@inaf.it}

\author[2,1]{\fnm{Mirko} \sur{Stumpo}}\email{mirko.stumpo@inaf.it}

\author[1]{\fnm{Giuseppe} \sur{Consolini}}\email{giuseppe.consolini@inaf.it}

\author[1]{\fnm{Tommaso} \sur{Alberti}}\email{tommaso.alberti@inaf.it}

\author[1]{\fnm{Monica} \sur{Laurenza}}\email{monica.laurenza@inaf.it}

\author[3]{\fnm{Emiliya} \sur{Yordanova}}\email{eya@irfu.se}

\affil*[1]{\orgdiv{INAF - Istituto di Astrofisica e Planetologia Spaziali}, \city{Roma}, \country{Italy}}

\affil[2]{\orgdiv{Dipartimento di Fisica}, \orgname{Università degli Studi di Roma Tor Vergata}, \city{Roma}, \country{Italy}}

\affil[3]{\orgname{Swedish Institute for Space Physics}, \city{Uppsala}, \country{Sweden}}

\abstract{In the framework of statistical time series analysis of complex dynamics we present a
multiscale characterization of solar wind turbulence in the near-Earth environment. The data analysis, based on the Markov-process theory, is meant to estimate the Kramers-Moyal coefficients associated with the measured magnetic field fluctuations. In fact, when the scale-to-scale dynamics can be successfully described as a Markov process, first- and second-order Kramers-Moyal coefficients provide a complete description of the dynamics in terms of Langevin stochastic process. The analysis is carried out by using high-resolution magnetic field measurements gathered by Cluster during a fast solar wind period on January 20, 2007. This analysis extends recent findings in the near-Sun environment with the aim of testing the universality of the Markovian nature of the magnetic field fluctuations in the sub-ion/kinetic domain.}




\maketitle

\section{Introduction}\label{sec1}

During last decades, high-resolution magnetic field measurements gathered in the inner Heliosphere have increased the interest in investigating the nature of the magnetic field fluctuations observed at scales below the ion inertial length \cite{Bruno16}. Observations gathered in the near-Earth environment were recently accompanied by high-resolution observations in the vicinity of the Sun, thus providing a comprehensive picture of the principal solar wind properties as well as their radial evolution \cite{chen2020evolution,Alberti20}. 
For interplanetary magnetic field (IMF) fluctuations observed at scales below 0.1-1 Hz, the physical phenomena show the universal features of fully developed turbulence with an important role played by the coupling between magnetic and velocity fields near the Sun and a fluid-like turbulence above $\sim0.6$ astronomical units. All these features are consistent with results obtained in the framework of fluid-like approximation of the solar wind turbulence from the integral scale of the energy injection to approximately $10\,d_i$, where $d_i$ is the ion inertial length \cite{Bruno16}.
As usually called in the literature \cite{leamon1998observational,kiyani2009global,chhiber2021subproton} from here on we will refer to this large scale interval characterized by a near-Kolmogorov spectrum as the \textit{inertial range}.
Conversely, there is no general consensus on what are the mechanisms responsible for the IMF fluctuations observed in the range below $d_i$, namely sub-ion/kinetic range. As a matter of fact, several mechanisms acting at sub-ion/kinetic scales have been proposed, such as wave-like fluctuations (e.g., Alfvén ion cyclotron, kinetic Alfvén waves, whistler waves), coherent magnetic structures (e.g., Alfvén vortices, current structures), magnetic reconnection processes \cite{boldyrev2013toward,lion2016coherent,gary2009,Schekochihin_2009,alexandrova2006,cerri2017reconnection}. Moreover, alternative approaches based on non-equilibrium statistical mechanics can also make important contributions in unveiling some characteristics of processes occurring at sub-ion/kinetic scales \cite{Carbone_2022}. An interesting method for investigating complex time series within the framework of stochastic processes has been developed and widely employed to characterize the energy transfer across the turbulent cascade \citep{pedrizzetti1994markov,friedrich1997description,friedrich1997statistical,renner2001experimental,peinke2019fokker}. These authors suggest that the turbulent cascade can be described as a Markov process whose two-points transition probabilities allow one to reproduce the evolution of the Probability Distribution Functions (PDFs) of the longitudinal velocity increments $u_\ell$ associated with the redistribution of energy across scales. More specifically, the amplitude of $u_\ell$ at subsequent scales within the inertial range can be represented as a Langevin process with a drift force and a diffusion strength that depend on the scale. In the field of space plasma turbulence, Strumik and Macek \citep{strumik2008statistical,strumik2008testing} shown that radial IMF fluctuations at typical time scales within the inertial range appear to satisfy the Markov property and the dynamics is accurately reproduced by means of the Fokker-Planck (FP) equation.

An important result by Kiyani \textit{et al.} \cite{kiyani2009global} using ESA-Cluster mission data, pointed out the existence of a global scale-invariance at scales below the ion-inertial length, thus suggesting an intriguing difference with respect to the anomalous scaling \citep[i.e., the deviation from the linear scaling predicted by the Kolmogorov theory,][]{Kolmogorov41}, which is ubiquitous in fully developed turbulence, being the fingerprint of intermittency. Analogous results were reported also in the case of the near-Sun environment through the high-resolution observations gathered by the Parker Solar Probe mission during the perihelia \citep[e.g., see][and references therein]{chhiber2021subproton}.
Regarding the Markov property in the sub-ion/kinetic range, Benella \textit{et al.} \citep{Benella_2022} recently showed that IMF fluctuations at these scales keep satisfy such condition in the near-Sun environment, thus sharing with the inertial range the local structure of the energy transfer across scales. However, the two regimes appear to be independent in a statistical sense, i.e. there is no correspondence between the IMF fluctuations measured in the inertial range and the corresponding fluctuations observed in the sub-ion/kinetic domain, thus supporting the idea that they are originated by different physical mechanisms. Since in this analysis we use a stochastic framework, it does not allow to discern between the different physical processes involved at sub-ion/kinetic scales, but only to characterize and to model the stochastic nature of the energy transfer, in terms of fluctuations, through different scales. This energy transfer is described in terms of an advection-diffusion stochastic equation through the scales. In fact, it has been shown that the evolution of the statistics of the IMF fluctuations within the sub-ion/kinetic range can be modeled through the FP equation. In this case, the observed global-scale invariance can be viewed as the stationary solution of the equation that governs the evolution in scale of the PDFs of the rescaled IMF fluctuations (i.e., IMF fluctuations divided by their standard deviations).

The aims of this work are two:
\begin{itemize}
    \item The first aim is to extend the analysis proposed in \citep{Benella_2022}, here referred as Kramers-Moyal (KM) analysis, to high-frequency near-Earth observations of solar wind gathered by the ESA-Cluster mission. In fact, the comparison between results obtained in different solar wind conditions at different heliocentric distances is crucial in order to look for an universality of these statistical properties at the sub-ion/kinetic scales;
    \item The second aim is to provide a general method for estimating the master curve of the rescaled IMF PDFs in the sub-ion/kinetic range by only using the KM coefficients evaluated on the IMF timeseries.
\end{itemize}

\section{Data and Methods}\label{sec2}

\subsection{Data}

\begin{figure}
    \centering
    \includegraphics[width=10cm]{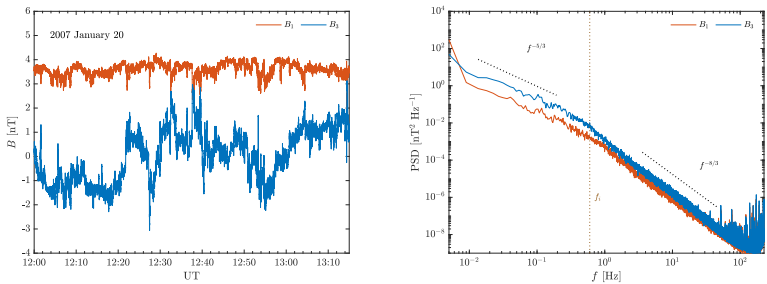}
    \caption{Left: IMF minimum (blue) and maximum (red) variance components on 20 January 2007 between 12:00 and 13:15. Right: PSD of $B_1$ (blue) and $B_3$ (red). Black dotted lines are $-5/3$ and $-8/3$ slopes, whereas the red dotted line indicates the ion-frequency $f_i$.}
    \label{fig:Fig1}
\end{figure}

The data we used in this study were gathered by the ESA-Cluster B spacecraft on 20 January 2007 in a 75 minutes time interval from 12:00 UT to 13:15 UT with a sampling rate of $\sim450$ samples/s. This period was characterized by a fast solar wind stream ($v\sim600$ km$^{-1}$) with a mean interplanetary magnetic field (IMF) of $B_0\sim4$ nT and a plasma density of $n\sim2$ cm$^{-3}$.
Since we are interested in IMF fluctuations parallel and perpendicular with respect to the mean field, we rotated the IMF components in the minimum variance reference frame. Therefore, we focus on the minimum ($B_1$) and maximum ($B_3$) variance components, which are mainly aligned along directions parallel and perpendicular to the mean field, respectively. The IMF components are reported in Figure \ref{fig:Fig1} (left panel) along with their power spectral densities (PSDs; right panel). These PSDs have an $f^{-\beta}$ behavior with two distinct regimes separated by a spectral break located around the ion-frequency $f_i\simeq 0.6$ Hz. For frequencies below $f_i$ we observe that the PSD exhibit a scaling \textit{\`a la} Kolmogorov (i.e., $\beta\sim-5/3$), whereas for frequencies larger than $f_i$ the spectral trend present the usual sub-ion/kinetic exponent $\beta\in[-7/3, -8/3]$ \citep{Bruno16,leamon1998observational,kiyani2009global}. In the following we perform the analysis in the time domain by assuming that we are exploring spatial scales $r$ via Taylor's hypothesis, i.e., $r=\tau \, U_0$, being $U_0$ the average plasma bulk speed.

\subsection{Methods}

The method we use in this work is based on the Markov process theory. By introducing the increments of the IMF $b_i(\tau)\doteq B(t+\tau)-B(t)$ (where $i=\{1,2,3\}$), we can interpret $b_i(\tau)$ as a stochastic processes that evolve across the time scales $\tau$. If the probability of observing the increment $b_{1,i}$ at the scale $\tau_1$ given the increments $b_{2,i}$ at the scale $\tau_2$ until $b_{\tau_n,i}$ at the scale $\tau_n$, with $\tau_1<\tau_2<\dots<\tau_n$ does not depend on increments at scales larger than $\tau_3$ the process satisfy the Markov condition, i.e.,
\begin{equation}
    p(b_{1,i},\tau_1\rvert b_{2,i},\tau_2;\dots;b_{n,i},\tau_n)=p(b_{1,i},\tau_1\rvert b_{2,i},\tau_2).
    \label{eq:markov}
\end{equation}
For a Markov process, the transition probability between $b_{1,i}$ and $b_{3,i}$, i.e., the probability of observing the fluctuation $b_{1,i}$ at scale $\tau_1$ given $b_{3,i}$ at scale $\tau_3$, can be written in terms of the Chapman-Kolmogorov (CK) equation by integrating over fluctuations $b_{2,i}$ at the intermediate scales $\tau_2$ such that $\tau_1<\tau_2<\tau_3$ \citep{risken1996fokker},
\begin{equation}
    p(b_{1,i},\tau_1\rvert b_{3,i},\tau_3)=\int_{-\infty}^{+\infty} p(b_{1,i},\tau_1\rvert b_{2,i},\tau_2)p(b_{2,i},\tau_2\rvert b_{3,i},\tau_3)db_{2,i}.
    \label{eq:ck}
\end{equation}
The time evolution of the transition probability is governed by the \textit{master equation}
\begin{equation}
   -\frac{\partial}{\partial \tau}p(b_i,\tau\rvert b_i',\tau')=\mathcal{L}_{KM}(b_i,\tau)p(b_i,\tau\rvert b_i',\tau'),
    \label{eq:master}
\end{equation}
where the operator $\mathcal{L}_{KM}(b_i)$ is the Kramers-Moyal (KM) operator and it is defined as
\begin{equation}
    \mathcal{L}_{KM}(b_i,\tau)=\sum_{k=1}^\infty\biggl(-\frac{\partial}{\partial b_i} \biggr)^k D^{(k)}(b_i,\tau).
\end{equation}
The functions $D^{(k)}(b_i,\tau)$ are called KM coefficients and are defined as
\begin{equation}
    D^{(k)}(b_i,\tau)=\frac{1}{k!}\,\lim_{\delta\to0}\,\frac{1}{\delta}\,\mathbb{E}[ (b_i(\tau-\delta)-b_i(\tau))^k\rvert b_i(\tau)]
    \label{eq:kmcoef}
\end{equation}
as also introduced in an early work by Kolmogorov \cite{kolmogorov1931analytical}.
The minus sign on the l.h.s. of Equation (\ref{eq:master}) is due to the direction of the time evolution from larger towards smaller scales \citep{renner2001experimental}.
In principle, the KM expansion encloses an infinite number of terms that contribute to the time evolution of the statistics of the process. An important class of stochastic processes is represented by those for which the Pawula's theorem is valid. In fact, the Pawula's theorem states that if $D^{(k)}(b_i,\tau)=0$, then all the coefficients of order $k\geq3$ vanish and the KM expansion reduces to the Fokker-Planck (FP) equation \citep{risken1996fokker}
\begin{equation}
  	 -\frac{\partial}{\partial \tau}p(b_i,\tau\rvert b_i',\tau')
   	 =\biggl[-\frac{\partial}{\partial b_i}D^{(1)}(b_i,\tau)+\frac{\partial^2}{\partial b_i^2}D^{(2)}(b_i,\tau)\biggr]p(b_i,\tau\rvert b_i',\tau').
    \label{eq:fp}
\end{equation}
The first-order KM coefficient $D^{(1)}(b_i,\tau)$ is representative of the \textit{drift} term, whereas the second-order KM coefficient $D^{(2)}(b_i,\tau)$ is related to the \textit{diffusion} strength, which modulates the amplitude of the delta-correlated Gaussian noise, $\Gamma(\tau)$, in the corresponding Langevin equation
\begin{equation}
    -\frac{\partial b_i}{\partial\tau}=D^{(1)}(b_i,\tau)+\sqrt{2D^{(2)}(b_i,\tau)}\,\Gamma(\tau).
\end{equation}
In this Langevin-like description of IMF fluctuations, the diffusion coefficient accounts for the stochastic character of the values assumed by the fluctuations at different scales. Since the true values of KM coefficients are not accessible from data, we will approximate them as their finite-time version for a sufficient small value of $\delta$
\begin{equation}
    D^{(k)}_{\delta}(b_i,\tau)=\frac{1}{k!\delta}\,\mathbb{E}[ (b_i(\tau-\delta)-b_i(\tau))^k\rvert b_i(\tau) ].
    \label{eq:km-res}
\end{equation}
Here, we applied the above technique to the small-scale increments of the magnetic field at the sub-ion/kinetic scales, starting from the verification of the CK Equation and successively evaluating the KM coefficients up to the fourth order.

\section{Results}\label{sec3}

\begin{figure}
    \centering
    \includegraphics[width=10cm]{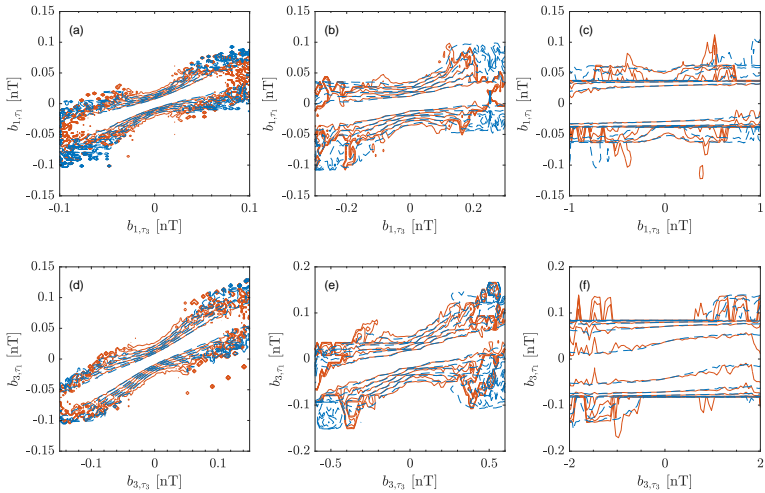}
    \caption{Comparison between $p_E$ (red curves) and $p_{CK}$ (blue curves) at different timescales for minimum (upper panels) and maximum (lower panels) variance directions. The timescale differences for the CK test are 0.02 s (panels a and d), 0.2 s (panels b and e), and 1.0 s (panels c and f).}
    \label{fig:Fig2}
\end{figure}

The first step of the analysis is to check the Markov property of $B_1$ and $B_3$ fluctuations at sub-ion/kinetic scales. By inspecting the PSD we observe that the instrumental noise affects magnetic field measurements around 100 Hz, thus we set the minimum time scale separation value to $\delta=0.02$ s. In order to test the validity of the Markov property we evaluate both the l.h.s. and the r.h.s. of the CK Equation (\ref{eq:ck}) and we compare the resulting transition probabilities. The l.h.s. of Equation (\ref{eq:ck}) represents the empirical transition probability $p_E$, whereas the r.h.s. is the CK transition probability $p_{CK}$. The results of the CK test are reported in Figure \ref{fig:Fig2} for three different values of the scale separation: $\tau_3-\tau_1=0.02$ s, $\tau_3-\tau_1=0.2$ s and $\tau_3-\tau_1=2$ s, being $\tau_1=0.02$ s. For sake of simplicity, the intermediate scale $\tau_2$ of the CK integral is chosen as the half of the scale separation, although the results are not affected by any particular choice of $\tau_2\in(\tau_1,\tau_3)$. Figures \ref{fig:Fig2}a and \ref{fig:Fig2}d show an excellent agreement between $p_E$ and $p_{CK}$ at 0.02 s scale separation for $b_1$ and $b_3$, respectively. A similar result can be observed at 0.2 s scale separation as reported in Figures \ref{fig:Fig2}b and \ref{fig:Fig2}e, thus pointing out that the magnetic field fluctuations $b_i(\tau)$ can be successfully approximated as a Markov process in scale until the smaller scales that are not affected by instrumental noise.
If the scale separation is pushed to 2 s, that corresponds to approach the low end of the inertial range, the CK equation is still valid, but values of the fluctuations $b_{i,\tau_1}$ observed at the smaller scale no longer depend on the correspondent values $b_{i,\tau_3}$ observed at larger scales. This result support the idea that IMF fluctuations observed in the sub-ion/kinetic and inertial ranges are statistically independent, being perfectly in agreement with previous findings in the near-Sun environment \citep{Benella_2022}.

\begin{figure}
    \centering
    \includegraphics[width=10cm]{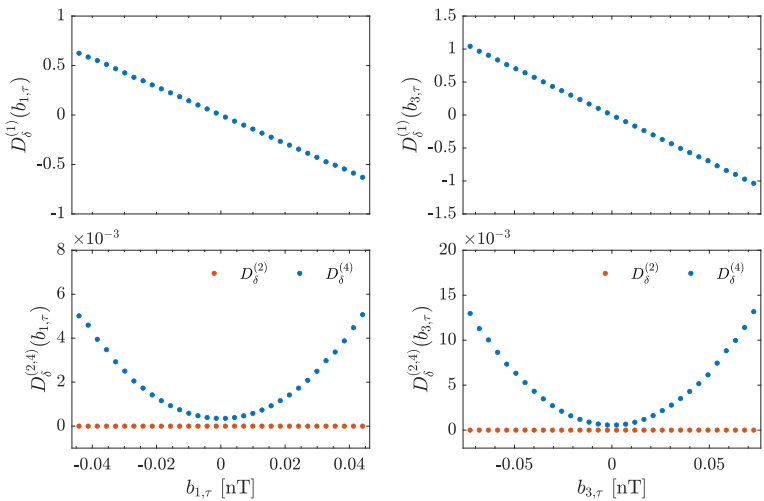}
    \caption{First-, second- and fourth-order finite-scale KM coefficients for minimum (left) and maximum (right) variance components of the IMF for $\tau=0.06$ s and $\delta=0.02$ s.}
    \label{fig:Fig3}
\end{figure}

The next step in the analysis is the evaluation of the first-, second- and fourth-order KM coefficients in order to check whether or not the Pawula's theorem is fulfilled. In Figure \ref{fig:Fig3} are reported the first-, second- and fourth-order KM coefficients for both minimum and maximum variance components with $\tau=0.06$ s and $\delta=0.02$. The fourth-order coefficient vanishes for $B_1$ and $B_3$ in the case of Figure \ref{fig:Fig3} (i.e., $\tau=0.06$ s) and for any $\tau$ in the sub-ion/kinetic range (not shown). This confirms that the evolution of the statistics of the process $b_i(\tau)$ as a function of $\tau$ is governed by the FP equation, whose drift and diffusion coefficients also depend on the scale.

One of the striking features of the sub-ion/kinetic range is the global-scale invariance which is generally shown by rescaling the PDFs by introducing the following transformation \citep{kiyani2009global,alberti2019multifractal,chhiber2021subproton}
\begin{align}
    b_{i,\tau}\rightarrow x_{i,\tau}=b_{i,\tau}/\sigma_{i,\tau}\\
    p(b_{i,\tau})\rightarrow p(x_{i,\tau})=\sigma_{i,\tau}\,p(b_{i,\tau}).
    \label{eq:btox}
\end{align}
The existence of a master curve, i.e., a PDF invariant in shape, can be interpreted in terms of stationary solution of the FP Equation (\ref{eq:fp}) across sub-ion/kinetic scales \citep{Benella_2022}.
We aim to derive such master curve starting from the KM coefficients evaluated for $b_{i,\tau}$. Since we are dealing with finite-time KM coefficients instead of their ``true'' value (i.e. taking the limit $\delta\to0$), the relation between the KM coefficients $D^{(k)}(b_i,\tau)$ and $D^{(k)}(x_i,\tau)$ is not straightforward. By substituting the transformation (\ref{eq:btox}) in Equation (\ref{eq:km-res}), we obtain the following relations between the KM coefficients
\begin{equation}
    D^{(1)}_\delta(x_i,\tau)=\frac{D^{(1)}_\delta(b_i,\tau)}{\sigma}+\\ \frac{1}{\delta}\biggl( \frac{1}{\sigma'}-\frac{1}{\sigma} \biggr)\mathbb{E}[ b_i(\tau-\delta)\rvert b_i(\tau) ],
    \label{eq:km-corr1}
\end{equation}
\begin{multline}
    D^{(2)}_\delta(x_i,\tau)=\frac{D^{(2)}_\delta(b_i,\tau)}{\sigma^2}+\frac{1}{2\delta}\biggl( \frac{1}{\sigma'^2}-\frac{1}{\sigma^2} \biggr)\mathbb{E}[ b_i^2(\tau-\delta)\rvert b_i(\tau) ]+ \\ - \frac{1}{\delta\sigma}\biggl( \frac{1}{\sigma'}-\frac{1}{\sigma} \biggr)\mathbb{E}[ b_i(\tau-\delta)b(\tau)\rvert b_i(\tau) ],
    \label{eq:km-corr2}
\end{multline}
where $\sigma$ and $\sigma'$ are the standard deviations of the IMF fluctuations at the scales $\tau$ and $\tau-\delta$, respectively. Starting from the calculation of all the finite-scale KM coefficients at different time scales within the sub-ion/kinetic range, Equations (\ref{eq:km-corr1}) and (\ref{eq:km-corr2}) enable us to evaluate all the corresponding finite-scale KM coefficients of the rescaled variables. By solving the FP equation for the stationary PDF, the general solution reads
\begin{equation}
    p_{ST}(x_i)=\mathcal{N}_i\exp\biggl\{-\ln D^{(2)}(x_i) +\int_{-\infty}^{x_i}\frac{D^{(1)}(\xi)}{D^{(2)}(\xi)}d\xi\biggr\}
    \label{eq:fp-steady}
\end{equation}
where $\mathcal{N}_i$ is the normalization factor.
\begin{figure}
    \centering
    \includegraphics[width=10cm]{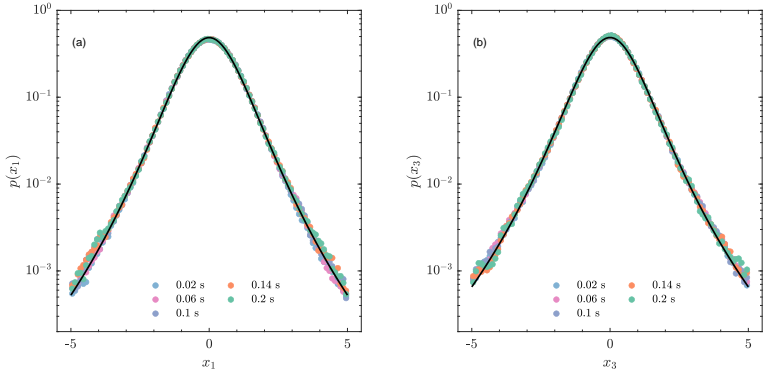}
    \caption{Comparison between the empirical PDF of the rescaled variable $x_i$ (circles) and the stationary solution of the FP equation (black lines) for the minimum (a) and maximum (b) variance components. The time scales used in the sub-ion/kinetic range range from 0.02 s to 0.2 s.}
    \label{fig:Fig4}
\end{figure}
A fundamental remark of our reasoning is that by evaluating the functions $D^{(1)}(x_i,\tau)$ and $D^{(2)}(x_i,\tau)$ through Equations (\ref{eq:km-corr1}) and (\ref{eq:km-corr2}) at any $\tau$ and by substituting them in Equation (\ref{eq:fp-steady}), the obtained stationary solution $p_{ST}(x_i)$ approximates the empirical master curve and does not depend on $\tau$. Figure \ref{fig:Fig4}ab shows the averaged function $p_{ST}(x_i)$ obtained through Equation (\ref{eq:fp-steady}) in the sub-ion/kinetic range (black lines) along with the empirical rescaled PDFs (circles). The stationary PDFs exhibit an excellent agreement with observations within $x_i\in[-5\sigma_{x_i}, 5\sigma_{x_i}]$ for both minimum and maximum variance components. We emphasize that the steady state solution has to be interpreted as the analogous of the \textit{PDF-invariant shape} observed in the sub-ion/kinetic range, which stems from the fact that the FP Equation (\ref{eq:fp}) describe the evolution of PDFs as a function of the time scale $\tau$. It has been shown that by parametrizing the KM coefficients as a linear ($D^{(1)}(x_i)$) and quadratic ($D^{(2)}(x_i)$) functions of $x_i$, the master curve of Figure \ref{fig:Fig4} can be written as a Kappa distribution \citep{Benella_2022}. In this work we have presented a more general derivation of the steady state solution in the rescaled variable $x_i$ that is based on the evaluation of KM coefficients on the variable $b_i$ at any $\tau$ the sub-ion/kinetic regime.

\section{Discussion and conclusions}\label{sec4}
A universal framework of the solar wind processes originating magnetic field fluctuations at sub-ion/kinetic scales is still lacking. Several mechanisms can contribute in generating IMF fluctuations at these scales, e.g., wave-like fluctuations, coherent magnetic structures, current patterns and magnetic reconnection \cite{boldyrev2013toward,lion2016coherent,gary2009,Schekochihin_2009,alexandrova2006,cerri2017reconnection}. In this regard the KM analysis of IMF fluctuations represents a novel approach to characterize the stochastic nature of physical processes acting at these scales. In this work we consider a fast solar wind stream in the near-Earth environment and we have given evidence that the description of the PDF evolution as a function of the time scale $\tau$ by means of Markov processes is successful at sub-ion/kinetic scales. Such property has been also verified for IMF fluctuations within the inertial range of magneto-fluid turbulence \citep{strumik2008statistical,strumik2008testing}, then supporting the idea that the energy flow from one scale to another has a local structure in both inertial and sub-ion/kinetic regimes. Although there is a net energy transfer between them, we provided evidence that a correspondence in terms of IMF fluctuations between inertial and sub-ion/kinetic ranges does not seem to emerge, confirming recent findings in the near-Sun environment \citep{Benella_2022}. This statistical independence between the observed fluctuations, together with the transition between different dynamical regimes observed in the PSD, supports the idea that the physical processes originating IMF fluctuations have a different nature in the two regimes. We remark that these are statistical results, therefore no direct information about the nature of the specific physical processes can be inferred, even though our results provide important constraints for models.

The main result reported in this work is represented by the clear Markovian nature of the IMF fluctuations in the sub-ion/kinetic range in the near-Earth solar wind. As pointed out in several recent works, in fact, many solar wind characteristics, i.e. IMF fluctuations, exhibit a radial dependence in terms of spectral features, intermittency, entropic character and so forth \cite{chen2020evolution,Alberti20,stumpo2021self}. By performing the KM analysis in the near-Earth IMF fluctuations we obtained similar results in comparison with the near-Sun solar wind conditions. This allows us to assert that such Markovian property of IMF fluctuations does not depend on the heliocentric distance and thus might represent a universal property in the sub-ion/kinetic domain.

The solar wind inertial range shows the typical anomalous scaling of fully developed turbulence, with an intermittent character of the energy transfer, whose strength depends on the heliocentric distance from the Sun \cite{Alberti20}. In terms of scale dynamics of the IMF fluctuation statistics, the intermittency is associated with a modification of the PDF across the inertial range. This process affects both IMF-fluctuation amplitude as well as the shape of their PDFs. Indeed, the higher probability of observing enhanced IMF fluctuations towards smaller scales is reflected in the modification of the Gaussian PDF at the integral scale into a leptokurtic function through the energy cascade. In the framework of Markov processes, the PDF dynamics across the inertial scales can be modeled through the general FP Equation, accounting for the variation of the shape of the PDF as a function of $\tau$. As a consequence, the absence of modification of the initial PDF across different scales can be due to the lack of intermittency in the physical processes accounting for the energy transfer. In this work we confirmed that the global scaling observed at sub-ion/kinetic scales can be interpreted by means of the stationary FP Equation (i.e., $\partial p/\partial\tau\to0$) since the PDFs of IMF fluctuations exhibit a well-defined shape-invariant form in this domain. 
From the physical point of view our results can be resembled in the following scenario:
\begin{itemize}
    \item as for the inertial range, also at sub-ion/kinetic scales the energy transfer is local in scale, i.e., it satisfies the Markov property, providing a complete description of the IMF fluctuations in terms of Langevin dynamics: the energy transfer turns out to be a stochastic process;
    \item the energy transfer mechanisms in the inertial range and in the sub-ion/kinetic domains originate IMF fluctuations exhibiting a statistical independence between the two regimes, thus they may have a different nature;
    \item the intermittency of IMF fluctuations observed in the inertial range is lost at sub-ion/kinetic scales, where a global scale invariance is observed.
\end{itemize}
A joint analysis involving both high-resolution magnetic and velocity field is desirable and will provide a more comprehensive picture of the statistics of the physical processes acting at sub-ion/kinetic scales.

\bmhead{Acknowledgments}

This work is funded by the Italian MIUR-PRIN grant 2017APKP7T on “Circumterrestrial Environment: Impact of Sun-Earth Interaction”. The authors acknowledge the Cluster FGM and STAFF P.I.s and teams and the ESA-Cluster Archive for making available the data used in this work. The FGM and STAFF data were combined using the IRFU-Matlab analysis package available at https://github.com/irfu/irfu-matlab. 
M.S. acknowledges the PhD course in “Astronomy, Astrophysics and Space Science” of the University of Rome “Sapienza”, University of Rome “Tor Vergata” and Italian National Institute for Astrophysics (INAF), Italy. E.Y. is supported by SNSA grants 86/20 and 145/18.

\bibliography{MarkovSwico.bib}

\end{document}